\documentclass{article}

\usepackage{graphicx}
\usepackage{epsfig}
\begin{document}

\hbox{} \nopagebreak \vspace{-3cm} \vspace{1in}

\begin{center}
{\Large \bf Perturbative Saturation and the Soft Pomeron}
\vspace{0.1in}

\vspace{0.5in}
{\large A. Kovner$^{a,b}$ and U.A. Wiedemann$^c$}\\
\vspace{.4in}
{\small
$^a$Department of Mathematics and Statistics,
University of Plymouth,\\
2 Kirkby place, Plymouth, PL4  8AA, UK\\
$^b$Department of Physics, University of Connecticut\\
Storrs CT 06269, USA\\
$^c$CERN, Theory Division, CH-1211
Geneva 23, Switzerland}\\
\vspace{0.5in}

{\bfseries\sc  Abstract}\\
\end{center}
\vspace{.2in} We show that perturbation theory provides two
distinct mechanisms for the power like growth of hadronic cross
sections at high energy . One, the leading BFKL effect is due to
the growth of the parton density, and is characterized by the
leading BFKL exponent $\omega$. The other mechanism is due to
the infrared diffusion, or the long range nature of the Coulomb
field of perturbatively massless gluons. When perturbative
saturation effects are taken into account, the first mechanism is
rendered ineffective but the second one persists. We suggest that
these two distinct mechanisms are responsible for the appearance
of two pomerons. The density growth effects are responsible for
the hard pomeron and manifest themselves in small systems (e.g.
$\gamma^*$ or small size fluctuations in the proton wave function)
where saturation effects are not important. The soft pomeron
is the manifestation of the exponential growth of the black saturated 
regions which appear in typical hadronic systems. 
We point out that the nonlinear generalization
of the  BFKL equation which takes into account wave function
saturation effects ("pomeron loops") provides a well defined
perturbative framework for the calculation of the soft pomeron
intercept. The conjecture of a perturbative soft pomeron is
consistent with picturing the proton as a loosely bound system
of several small black regions corresponding e.g. to constituent 
quarks of size about 0.3 fm.
Phenomenological implications of this picture are
compatible with the main qualitative features
of data on p-p scattering. \vfill

\newpage
\section{Introduction}
It has been long recognized that at
asymptotically high energies hadronic cross sections are dominated by
soft nonperturbative physics. In particular the validity of the
Froissart bound for the total cross section requires a mass gap 
in the spectrum. The corresponding generation of the pion mass in
QCD, or the generation of the glueball mass in pure gluodynamics,
is a bona fide nonperturbative effect. However, even though the 
asymptotics is expected to be nonperturbative, perturbative 
dynamics may well play an important role in the preasymptotic 
regime.

Indeed, cross sections for small objects, like highly
virtual photon $\gamma^*$, or heavy quarkonium ("onium") are
perturbative up to very high energies, as long as the size of the
system remains small. The main perturbative mechanism that drives
the growth of cross sections with energy is the BFKL evolution
\cite{bfkl}. It predicts that cross sections grow
exponentially, $\sigma\propto s^{\omega}$, where $\omega=4\ln
2N_c{\alpha_s\over\pi}$  to leading order in $\alpha_s$. This
behaviour is nonunitary, as it violates the Froissart bound
$\sigma\le{\pi\over m^2}t^2$. Here $t$ is the rapidity,
$t=\ln{s\over m^2}$, and $m$ denotes the mass of the lightest particle
in the theory. The main reason for this lack of unitarity is the
growth of the partonic density in the hadronic system. As the
partonic density reaches a critical value of order $1/\alpha_s$,
the BFKL approach ceases to be valid, and one must take into
account finite density effects.

This has been recognized and forcefully advocated in the
pioneering works \cite{glr} and later in \cite{mv}. 
A more complete approach to 
QCD evolution at finite density has been developed more recently 
in \cite{jklw}. These papers treat the QCD evolution to leading 
logarithmic approximation, but to all orders in the gluon density. 
The resulting system of evolution equations is a 
set of functional equations, and its study so far has not 
been feasible beyond the double logarithmic approximation.

There exists however a regime in which the general nonlinear evolution 
simplifies. This happens when a small object (such as 
highly virtual photon) scatters on a large target (such as a
large nucleus). In this case the leading nonlinear corrections  
are due to the fact that the projectile wave function at 
high energy has a large multigluon component. As soon as the
{\it number} of gluons in these multigluon states becomes large, 
one has to account for the possibility that more than one gluon scatters,
even if the gluonic {\it density} may still be small.
When the target is large, so
that the scattering probability of a gluon is parametrically larger 
than $\alpha_s$ (for a large nucleus of atomic number $A$, it is 
$O(A^{1/3}\alpha_s)$), these corrections become important earlier than 
those due to high density effects.

The system of evolution equations which takes into account these
multiple scattering effects has been derived first in
\cite{balitsky}. In \cite{kovchegov} its large $N_c$ limit was
derived independently in the dipole picture of \cite{muellerdip}.
The advantage of the large $N_c$ limit is that the (otherwise
infinite) hierarchy of evolution equations closes and becomes a
single equation for the gluon density (or dipole scattering
probability). We will refer to this nonlinear equation as the BK
equation. The relation between the equations of \cite{jklw} and
\cite{balitsky} has been discussed in \cite{agh}, where it was
shown that the latter is the limit of the former when the induced
field density is small. This result was later rederived by similar
methods in \cite{ael}\footnote{ We note that although it is
claimed verbally in \cite{ael} that the equations of \cite{jklw}
should be equivalent to those of \cite{balitsky} not only for
small induced fields, but also in general, the actual mathematical
analysis of \cite{ael} does not justify this claim. The
mathematical analysis of \cite{ael} (up to notational differences)
is equivalent to that of \cite{agh}, where the origin of the
differences between the results of \cite{balitsky} and \cite{jklw}
has been discussed.}.

The range of validity of the BK equation is wider than that of the
BFKL evolution in the sense that it can be applied to
scattering on large targets. If the target is large so that the
scattering probability of a given probe on it is of order unity,
the BFKL evolution violates unitarity very quickly, at rapidities
of order $t_{BFKL}\propto 1/\alpha_sN_c$. This violation of unitarity
stems from the scattering probability rising above one locally
in impact parameter space. In contrast to BFKL, the BK evolution 
ensures that the scattering probability at a given impact parameter 
is always below unity. Nevertheless, as we will discuss in the
following, for scattering of a (small) projectile of transverse
size $x_0$ on a (large) target of size $R_0$, the BK evolution
violates unitarity for rapidities $t>t_{BK}\propto 1/\alpha_sN_c\ln
(R_0/x_0)$. The reason for this violation is that the value of the
maximal impact parameter that contributes to scattering grows
exponentially with rapidity.  Thus even though the scattering
probabilities remain unitary, the total cross section grows exponentially, 
$\sigma\propto\exp\{\epsilon t\}$, and violates the
Froissart bound. This is the
main result of \cite{sadly}. We can give a crude estimate of the
exponent $\epsilon$, but we are not able to perform a reliable
analytical calculation at this point. However the numerical
results of \cite{salam} give $\epsilon=0.75\omega$, which is
indeed compatible with our rough estimate. Thus the BK exponent is
smaller than the BFKL one, but not very significantly.

The reason why at high enough energy the BK evolution ceases to be 
valid is that the projectile wave function becomes dense, and further 
evolution is affected by the high density effects in the projectile wave
function. The onset of these corrections is at $t\propto{1\over
\alpha_sN_c}\ln{1\over \alpha_s}$. As discussed below, the gluon density 
in the center of the projectile is large at
energies at which large values of the impact parameter become
dominant.
Thus, finite density corrections in the central region
are important. Their main direct effect is to slow down the
growth of density in the central region, but this must also have a
large effect on the growth of the scattering probability at large
impact parameters. As we will discuss below, this growth 
at large impact parameter is due to
the long range Coulomb (or Weizs\"acker-Williams) fields originating
from the fluctuation of the colour charge in the central "black"
region. The finite density effects inevitably reduce the magnitude
of these fluctuations and therefore also the Coulomb fields which
feed the periphery. We expect therefore that these corrections
 are likely to reduce the exponential $\epsilon$
significantly. The proper calculation of the exponential
$\epsilon$ must involve the analysis of the full nonlinear
equation, including the wave function saturation effects \cite{jklw}.

One aim of the present paper is to furnish more details
to the derivations of \cite{sadly}. This is
done in sections 3 and 4, after the short recourse to the BFKL
evolution in section 2. The other aim is to reflect on possible
implications of these two distinct perturbative
mechanisms for the power like growth of the total cross section.
The first mechanism - the growth of partonic density - is naturally
associated with small systems. As long as the size of the system
is small and also the number of partons not very large, one
expects that the growth of density is indeed the leading
mechanism. This suggests that the BFKL exponent
(possibly modified by higher order corrections) dominates the
scattering cross section of small systems. However, if
a typical hadron is a dense partonic system rather than a
dilute one, then the density growth mechanism is rendered irrelevant
by saturation effects. It is only the power like growth due to the 
perturbative expansion in the transverse plane which remains effective 
in this case. One is thus lead naturally to conjecture, that it 
is this latter mechanism that is responsible for the experimentally 
found "soft pomeron" behaviour of hadronic cross sections.

Associating the soft pomeron with a perturbative phenomenon is 
certainly unconventional. All models of the soft pomeron that we 
are aware of, try to explain the power growth of cross sections by a
nonperturbative mechanism \cite{kopelovich,dima}. On the other
hand, this power growth of the cross section implies by the
extension of Heisenberg's argument \cite{heisenberg} the power
like distribution of "matter" in impact parameter space. 
Perturbation theory naturally provides such a mechanism, since it operates 
with massless gluons and thus long range Coulomb fields. 
Thus it could well be
that the role of nonperturbative physics is limited to 
taming the powerlike perturbative growth in asymptotia, rather 
than to provide an additional mechanism for a power like
growth in the pre-asymptotic regime!\footnote{ We note that the
models of \cite{kopelovich,dima}, when translated into our
language try to provide an alternative nonperturbative mechanism
for the growth of the density rather than for the growth in impact
parameter space. Since the exponent governing this growth is supposed
to be much smaller than the perturbative one, it is hard to see
how this sort of mechanism can survive as the leading one when
juxtaposed with BFKL.}

For this scenario to be viable, the basic saturated ("black") 
building blocks of hadrons must be themselves small in
size, so that the perturbative Coulomb like structure of the gluon
fields is still relevant at these distances. It has been pointed out 
repeatedly in the literature that the distance scale associated with the
constituent quarks is much smaller than
the scale of confining physics. Indeed, the models 
\cite{kopelovich,dima} deal with scales of order .3 fermi. The
picture we suggest therefore,
is that of a proton containing
three small "black" constituent quarks with the size about .3 fermi in the
proton rest frame. As the proton is boosted, the radius of
constituent quarks grows with half the soft pomeron power, 
$r\propto s^{.04}$. The mechanism of this
growth is {\it perturbative} and the power itself should be
calculable from the full nonlinear QCD evolution equation
\cite{jklw}. Section 5 of the present paper is devoted to a more
detailed discussion of this scenario.

\section{The two exponents of the BFKL equation}

To illustrate the perturbative features of high energy reactions,
we choose the example of Deeply Inelastic Scattering. The subprocess 
relevant to strong interaction physics is the scattering of the photon of
virtuality $Q^2$ on a proton. In the parton model, the photon counts 
the number of charged partons in the proton. The scattering cross 
section is then
\begin{equation}
\sigma_{DIS}(Q,x)={\alpha_{em}\over Q^2}\sum_ie_i^2N_i(Q,x)\, ,
\label{DIS}
\end{equation}
where the first factor is the parton level cross section while
$N_i$ is the number of partons of a given species in the proton.
The number of partons depends on both, the resolution scale $Q^2$
and the energy at which the proton wave function is probed,
$x={Q^2\over Q^2+W^2}$. Here $W$ denotes the center of mass energy of
the $\gamma^*p$ system. One can define the phase space density of
partons, $\phi(k)$,  in terms of which
\begin{equation}
  N(Q^2,x)=\int_S d^2b\int_{k^2\le Q^2}d^2k\, \phi(b,k,x)\, ,
\end{equation}
where $k$ is the intrinsic transverse momentum of the parton, and
$b$ is the impact parameter at which the parton is found in the
transverse plane.

Although the photon directly counts the number of charged partons,
i.e. quarks, at low $x$ the proton wave function is dominated by
gluons. The number of quarks is then directly determined by the
gluon content of the wave function. In the following we will
therefore concentrate on the gluons only.

At low $x$, in the leading logarithmic approximation, the gluon
density $\phi$ is determined by the asymptotic solution of the
BFKL equation\cite{bfkl}. According to this solution, the
distribution of gluons is
\begin{equation}
  \phi(b,k,x,k_0)\propto  \exp\{\omega t -{\ln^2 {b^2kk_0}\over
    a^2t}\}\, . \label{gluons}
\end{equation}
Here $t=\ln 1/x$ is the rapidity, $\omega=4\ln 2 N_c\alpha_s/\pi$
and $a^2=14\zeta(3)N_c\alpha_s/\pi$ with $\zeta(n)$ being the Riemann
zeta-function, $\zeta(3) = 1.202\dots$. This distribution depends 
also on $k_0$ which
characterizes the initial transverse momentum of the gluon which
gave rise to $\phi$ through the evolution to high $t$.
Formula (\ref{gluons}) is valid for impact parameters which are 
not too large,
namely
\begin{equation}
  0\le\ln{b^2kk_0} \ll \alpha_s t\, .
\end{equation}

The striking feature of eq.(\ref{gluons}) is that even if
at low rapidity $t_0$ one starts with a single gluon,
after evolution to high enough $t$, the density of gluons at small
impact parameters becomes exponentially large. Within the
diffusion radius $\ln b\propto at^{1/2}$, the overall scale of
density  is determined by the exponential factor $\exp\{\omega
t\}$. Thus using eq.(\ref{gluons}) in the gluonic analog of
eq.(\ref{DIS}) gives the total cross section which grows
exponentially with rapidity.

The high density at central impact parameters is not the only 
source of growth in eq.(\ref{gluons}). The gluon density 
does decrease towards the peripheral impact parameters, but 
it does so rather slowly.
For a given transverse momentum $k$, the
density stays finite up to impact parameters of order
$\ln(b^2kk_0)\propto \alpha_s t$. One can not establish the exact
proportionality coefficient from eq.(\ref{gluons}), since the
validity of this equation does not extend to such large impact
parameters. However impact parameters of order $\ln(b^2kk_0)=\nu
\alpha_s t$ with very small $\nu$ are still covered by
eq.(\ref{gluons}). For such impact parameters the density
is still exponentially large, although parametrically smaller than
inside the "diffusion radius". These peripheral impact parameters
do not contribute to the leading BFKL exponential.
When the density is integrated over the
impact parameter plane to calculate the total number of gluons,
one obtains the leading BFKL result $N(t)=\int
d^2b\, \phi(b,t)=a\exp{\omega t}$ which is dominated by the impact
parameters within the diffusion radius. However the contribution
of the peripheral region is by itself also exponentially
increasing. Were we to exclude the central impact parameters from
the integration region, we would still get an exponentially large
contribution of the form $\exp\{\epsilon\alpha_s t\}$. The exponent 
here is smaller than the leading BFKL one, and thus gives a negligible 
contribution to the total cross section within the BFKL
framework.  However, this exponent is nevertheless 
present, and its physical origin is quite distinct
from the exponential growth of partonic density.

The exponential growth of the cross section 
violates the Froissart bound which
requires the total cross section to grow not faster than the
second power of the logarithm of energy
\begin{equation}
\sigma_{total}\le{\pi\over m_\pi^2}\ln^2{s\over m_\pi^2}\, .
\end{equation}

The leading exponential growth of the BFKL cross section is of
course unphysical at high enough energies. 
Taken at its face value
eq.(\ref{DIS}) in conjunction with eq.(\ref{gluons}) would mean
that the probability for scattering of a strongly interacting
probe at fixed impact parameter grows without bound at large
energy. This is inconsistent with the fact that the
probability can not exceed one, and violates unitarity of the
scattering probability at fixed impact parameter. The reason is 
that a strongly interacting probe has finite probability to 
interact with more than one gluon at fixed $b$ when the gluon 
density at fixed impact parameter and fixed resolution $Q^2$ 
exceeds $1/\alpha_s$. Thus the cross section is not
proportional to the number of gluons anymore. Multiple scattering
effects must be properly taken into account in order to relate the
gluon density with the scattering probability.

A nonlinear QCD evolution equation that takes into account these
multiple scattering effects has been derived by Balitsky
\cite{balitsky}. Its large $N_c$ version was obtained by
Kovchegov \cite{kovchegov} using Mueller's dipole model approach.
In the next two sections we will discuss how far this perturbative
resummation goes beyond the simple BFKL framework towards
restoring the unitarity of the hadronic cross sections.

We start our discussion with considering the evolution in the
target rest frame.
%
\begin{figure}[h]\epsfxsize=11.7cm
\centerline{\epsfbox{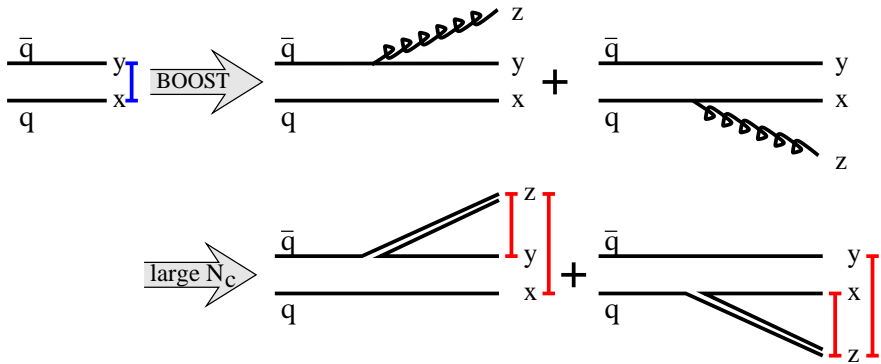}}
\caption{Boosting the $q\bar{q}$-dipole $(x,y)$ generates the 
higher Fock component $|q(x)\, \bar q(y)\, g(z)\rangle$ with a gluon
at transverse position $z$. In the large $N_c$-limit, this 
corresponds to the generation of dipoles $(x,z)$ and
$(y,z)$.
}\label{fig1}
\end{figure}
%
\section{The BK evolution in the target rest frame}
The essence of the BK evolution is the following. Suppose at some
initial rapidity $t_0$ we are interested in the scattering of a
probe consisting of a $\bar qq$ dipole on a large hadronic target,
for example a heavy nucleus. The scattering probability 
of a dipole with legs at transverse coordinates $x$ and $y$ on a 
hadronic target is $N(x,y)$. Increasing the center of mass energy 
(or rapidity) amounts to boosting the dipole to rapidity
$t=t_0+\delta t$. Under boost, the longitudinal Coulomb field 
associated with the dipole acquires a transverse part, i.e., the
dipole of transverse size $x-y$ generates an extra gluonic component
whose density is given by the equivalent gluon content of the 
Weizs\"acker-Williams field of the dipole, see Fig.~\ref{fig1}:
\begin{equation}
|q(x)\, \bar q(y)\rangle \rightarrow  A|q(x)\, \bar q(y)\, g(z)\rangle
\end{equation}
with
\begin{equation}
A^2={\alpha_sN_c\over 2\pi^2} \delta t{(x-y)^2\over
(x-z)^2(x-y)^2}\, .
\end{equation}
In the large $N_c$ limit, the gluon is equivalent to a quark-antiquark 
pair. This and the global colour conservation of the QCD evolution 
implies that the singlet $\bar q q g$ state is equivalent to two 
dipoles with coordinates $x,z$ and $y,z$ respectively, see Fig.~\ref{fig1}. 
In the leading
large $N_c$ approximation, the dipoles scatter independently of
each other. Thus the probability for the scattering of the pair of
dipoles is
\begin{equation}
N(x,z;y,z)=N(x,z)+N(y,z)-N(x,z)N(y,z)\, .
\label{ps}
\end{equation}

The last negative term, $N(x,z)N(y,z)$,
is the probability 
that {\it both} dipoles 
scatter in the same collision. Such double scattering events should 
be counted once and not twice in the total cross section, and the last term in 
eq. (\ref{ps})
corrects the overcounting of $N(x,z)+N(y,z)$.

This leads to the following nonlinear evolution equation for the
dipole scattering probability:
\begin{eqnarray}
  \label{kov}
  &&{d\over dt}N(x,y) = {\alpha_sN_c\over 2\pi^2}\int
                        d^2z{(x-y)^2\over(x-z)^2(y-z)^2}
   \\
&&\qquad  [N(x,z)+N(y,z)-N(x,z)N(z,y)-N(x,y)] \, .\nonumber
\end{eqnarray}
The first three terms in eq. (\ref{kov}) are just the ones discussed
above, while the last term is the "virtual" correction which
ensures that the dipole wave function stays normalized throughout
the evolution.

The following two properties of the BK equation eq.(\ref{kov}) are
important for our discussion. First, as stressed above, it takes
into account multiple scattering corrections. Second, within this
approach  the projectile wave function evolves according to the
linear evolution of the dipole model. Interactions between the
dipoles in the projectile wavefunction are not taken into account.
This is seen explicitly e.g. from the original
derivation of \cite{kovchegov} where 
the density of dipoles in the projectile wave function $n_1(x,y)$
satisfies the linear BFKL evolution equation,
\begin{eqnarray}
  \label{bfkl}
  &&{d\over dt}n_1(x,y) = {\alpha_sN_c\over 2\pi^2}\int
                        d^2z\\
     &&[{1\over(y-z)^2}n_1(x,z)+{1\over(x-z)^2}n_1(y,z)-
       {(x-y)^2\over(x-z)^2(y-z)^2}n_1(x,y)]\, .\nonumber
\end{eqnarray}
The nonlinearity in the BK evolution equation 
comes not from the nonlinearities in the evolution of
the projectile wave function, but from the nonlinearity
in the relation between the dipole density and the scattering
probability. This is again given explicitly in \cite{kovchegov}. A
single dipole $(x_0,y_0)$ at initial rapidity $t_0$ develops at a
greater rapidity $t$ into a wave function characterized by the
$m$-dipole densities, $n_m(x_0,y_0,t_0|x_1,y_1;...;x_m,y_m,t)$. 
If the single dipole scattering probability at $t_0$ is
$\gamma(x,y)$, the total scattering probability is given 
by~\cite{kovchegov}
\begin{equation}
  N(x_0,y_0,t)=\sum_m[\Pi_{i=1}^m\gamma(x_i,y_i)]\, 
  n_m(x_0,y_0,t_0|x_1,y_1;...;x_m,y_m)\, .
\end{equation}
The scattering probability $\gamma$ depends on the
target, but not on the rapidity.

The account of multiple scatterings in the BK resummation
eliminates the leading mechanism that renders the BFKL evolution
nonunitary. Since the scattering probability in the BK evolution
is no longer proportional to the gluon density, the scattering
probability at any impact parameter does not exceed unity. This 
is obvious from eq.(\ref{kov}). When the scattering probability 
$N(x,y)$ reaches unity at all impact parameters, the right hand 
side of the evolution equation vanishes, and the probability stops growing.

A number of numerical
\cite{lt}\cite{braun}\cite{gms} as well as analytical
\cite{kov1},\cite{el} studies of eq.(\ref{kov}) have been performed,
and they all lead to the following consistent picture: Suppose one 
starts the evolution
from the initial condition of small target fields (or $N(x,y)\ll 1$
for all $x,y$). Then initially the evolution follows the BFKL
equation, since the nonlinear term in eq. (\ref{kov}) is
negligible. As the scattering probability approaches unity, the
nonlinear term kicks in and eventually the growth stops as the RHS
of eq.(\ref{kov}) vanishes for $N(x,y)=1$. The larger dipoles
(large $(x-y)^2$) saturate earlier, with the smaller dipoles
following at later "time" $t$. The following simple parametrization
\cite{gw} of the scattering probability gives an adequate
description of the evolution
\begin{equation}
  N(x,y)=1-\exp\{-(x-y)^2Q^2_s(t)\}
  \label{golec}
\end{equation}
with the saturation momentum $Q_s(t)$ a growing function of
rapidity. Thus at any given value of rapidity, all pairs of size
greater than $Q_s^{-1}(t)$ are saturated.

The exact dependence of $Q_s$ on rapidity is not known, but both,
the numerical results \cite{lt},\cite{gms} and simple theoretical estimates
\cite{mueller},\cite{el} are consistent with the exponential
growth of the form
\begin{equation}
   Q_s(t)=\Lambda\exp\{\alpha_s\lambda t\} \label{qs}
\end{equation}
with $\lambda$ of order unity. This physical picture has been
anticipated several years ago in \cite{mueller}.

Does saturation of the scattering probability {\it locally} in 
impact parameter plane necessarily imply that the total cross
section unitarizes and satisfies the Froissart bound?
The answer clearly is negative.
The Froissart bound states that the 
inelastic cross section for the scattering of a hadron (dipole) on
a hadronic target can not grow faster than the square of rapidity
\begin{equation} 
  \sigma<{\pi\over m^2} t^2\, ,
\end{equation}
where $m$ is the mass of the lightest hadronic excitation. To
calculate the inelastic cross section one has to integrate
the scattering probability over the impact parameter. Thus, in the
saturation regime
\begin{equation}
  \sigma=\pi R^2(t)\, ,
\end{equation}
where $R(t)$ is the size of the region in the transverse plane in
which the scattering probability for hadronic size "dipoles" is
unity. This radius itself depends on $t$. To satisfy the Froissart
bound the radius $R(t)$ should grow at most linearly with $t$. The
question of unitarity is therefore the question about the rate of
growth of the "black" region, and thus is completely separate
from the
question of saturation of the scattering probability at fixed
impact parameter. 

As an aside we note that for the Deeply Inelastic Scattering the
unitarity bound is somewhat different. In this case the projectile
is a virtual photon. It does not have a fixed hadronic size, but
rather is characterized by a distribution of dipole sizes. The
perturbative wave function of the virtual photon is well known. An
interesting property of this wave function is that for transverse
photon it has a long tail on the small dipole side ($r^2<<Q^{-2}$)
\begin{equation}
  \Phi^2(r)\propto \alpha_{em}{1\over r^2}\, .
\end{equation}
In such a projectile not all dipoles saturate at the same energy.
The scattering probability thus is given by the integral over the
dipole sizes. At every rapidity the main contribution comes from
the dipoles which are saturated, that is those with sizes above
$Q_s^{-1}(t)$. The scattering probability for a virtual photon at
high energy (by high we mean here such that $Q_s>>Q$), is given by
\begin{equation}
  N(\gamma^*)=\int_{r^2<Q_s^{-2}} d^2r
  \Phi^2(r)\propto\alpha_{em}\ln{Q_s/Q}\, .
\end{equation}
With the exponential dependence of $Q_s$ on rapidity this
translates into\footnote{Although this is entirely academic, we
note that this expression is valid only at energies below $s_c$
such that $\alpha_{em}\alpha_s \ln s_c/s_0<1$. Above this energy
higher order electromagnetic corrections must become important so
that $N(\gamma^*)$ saturates.}
\begin{equation}
N(\gamma^*)\propto\alpha_{em}\alpha_s \ln s/s_0\, ,
\end{equation}
and therefore
\begin{equation}
\sigma_{DIS}\propto\alpha_{em}\alpha_s\pi R^2(t)t\, .
\end{equation}
Thus the DIS cross section has an extra power of $t$ relative to
the cross section of a purely hadronic process. This extra power
of $t$ is consistent with the numerical results of \cite{gms}.
The basic question of unitarization however remains the same: what is
the dependence of $R$ on rapidity ?

While there is no doubt that QCD is a unitary theory, and
therefore indeed $R(t)\propto t$, there is no guarantee that the
nonlinear BK equation eqs.(\ref{bal}),(\ref{kov}) preserves this
property. In fact simple considerations indicate the opposite.
This is especially clear from Kovchegov's derivation \cite{kovchegov} (see 
also \cite{ua}) where the density of dipoles in the projectile wavefunction
is explicitly determined by the BFKL equation. 
Saturation is the result of the multiple scattering of the dense
dipole system, rather than the slowdown in the growth of the
dipole density. Since the transverse size of a system in BFKL
evolution grows exponentially with rapidity,
there is little doubt that the BK evolution violates unitarity
of the total cross section.
%
\begin{figure}[h]\epsfxsize=10.7cm
\centerline{\epsfbox{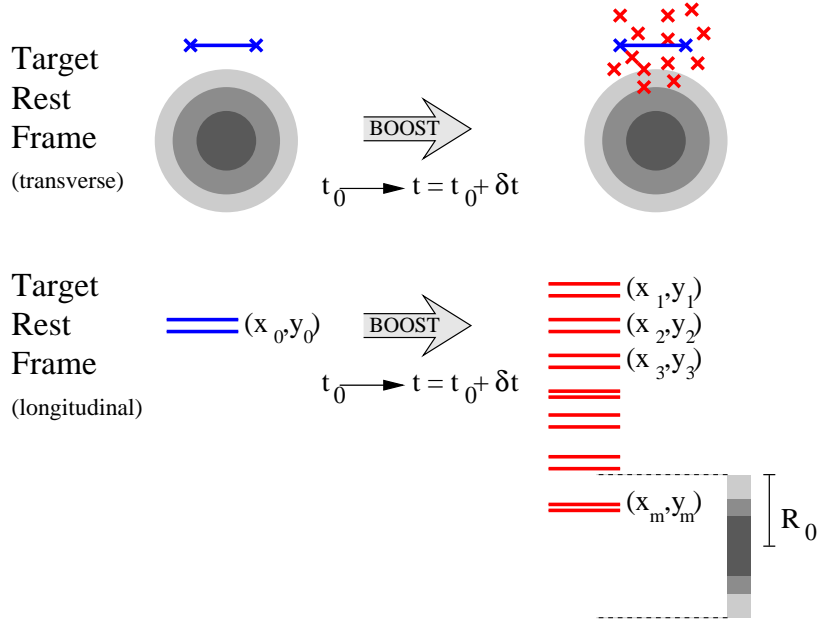}}
\caption{The BK evolution in the target
rest frame. Under boost, the initial dipole $(x_0,y_0)$ evolves
into a wavefunction containing $m$-dipole configurations with
density determined by the BFKL expression (\ref{density}). 
The spread of these configurations in impact parameter space 
leads to a finite interaction probability even if the initial
dipole $(x_0,y_0)$ was at very large impact parameter.
}\label{fig2}
\end{figure}

We now present a simple calculation that establishes this point:

Consider the BK evolution as the evolution of the projectile
\cite{kovchegov},\cite{ua}. Suppose at the initial energy the
projectile is a colour dipole of size $x_0$. It scatters on a
hadronic target of size $R_0$. As the energy is increased, the
projectile wave function evolves according to the BFKL equation.
At rapidity $t$ the density of dipoles of size $x$ at transverse
distance $r$ from the original dipole is given by the BFKL
expression (see for example \cite{forshaw}):
\begin{equation}
n(x_0,x,b,t)={32\over x^2}{\ln {16b^2\over x_0 x}\over(\pi
a^2t)^{3/2}} \exp\{\omega t-\ln {16b^2\over x_0 x} -{\ln^2
{16b^2\over x_0 x}\over a^2t}\}
\label{density}
\end{equation}
with $\omega=4\ln 2 N_c\alpha_s/\pi$ and
$a^2=14\zeta(3)N_c\alpha_s/\pi$ and $\zeta(n)$ being the Riemann
zeta-function.

\begin{figure}[h]\epsfxsize=11.7cm
\centerline{\epsfbox{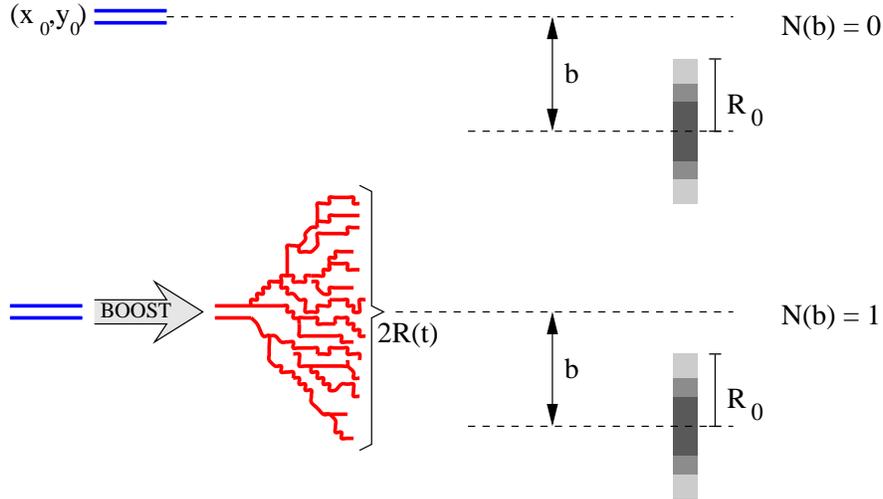}}
\caption{Under boost, the dipole density of the BK equation evolves
according to (\ref{density}). If the density of dipoles of 
critical size $Q_s^{-1}(t_0)$ is unity at some impact parameter 
$b$, then the scattering probability $N(b)$ is unity as well.
}\label{fig3}
\end{figure}

Once the density of dipoles at some impact parameter $b$ becomes
larger than some fixed critical number, the scattering probability
at this impact parameter saturates. The exact value of this number
depends on the target, but importantly it does not depend on
rapidity. Thus the total cross section is given by the square of
the largest impact parameter at which the dipole density in the
projectile wave function is of order unity. In order to estimate
this directly from eq. (\ref{density}), we choose the dipole size
$x$ in (\ref{density}) as $x=Q_s^{-1}(t_0)$. Recall that according
to eq.(\ref{golec}), the dipole of this size scatters with
probability one, if it hits inside the radius of the target $R_0$
(in this view of the evolution only the projectile wave function
depends on energy, while the properties of the target at $t$ are
the same as at $t_0$). Thus if at some impact parameter $R(t)$ the
density of dipoles of size $Q_s^{-1}(t_0)$ is unity, the
scattering probability at this impact parameter is unity as well,
see Fig.~\ref{fig3}.
Requiring the exponential in eq.(\ref{density}) to vanish we
obtain \cite{expon}
\begin{equation}
  R^2(t)={1\over 16}{x_0\over Q_s(t_0)}\exp\{{\alpha_sN_c\over 2\pi}
          \epsilon t\} \label{radius1}
\end{equation}
with
\begin{equation}
  {\alpha_sN_c\over 2\pi}\epsilon={a^2\over
   2}[-1+\sqrt{1+4{\omega\over a^2}}]\, . \label{epsilon}
\end{equation}
Numerically we find
\begin{equation}
  {\alpha_sN_c\over 2\pi}\epsilon=.87 \omega\, . \label{epsilon1}
\end{equation}
 Thus, as claimed we arrive at the exponential
growth of the total cross section.
%

The exact value of $\epsilon$ given in
eqs.(\ref{epsilon},\ref{epsilon1}) should not be taken too
seriously. The point is that the explicit form of the dipole
density eq.(\ref{density}) was derived by a saddle point
integration, and as such is valid only for $\ln {16b^2\over x_0
x}<\alpha_st$. Since this condition is not satisfied by
eq.(\ref{radius1}), we can not strictly speaking use the saddle
point expression eq. (\ref{density}). This ambiguity however
affects only the numerical value of $\epsilon$ and not the
parametric dependence in eq. (\ref{radius1}). The reason is that
even beyond the saddle point approximation the density has the
form
\begin{equation}
  n(x_0,x,b,t)\propto{1\over x^2} \exp\{\alpha_s t F({\ln
  {16b^2\over x_0 x}\over \alpha_s t})\}\, . \label{density1}
\end{equation}
The relevant condition is $F=0$, and thus the solution
parametrically must be the same as eq. (\ref{radius1}).

It is important to realize, that although we use the BFKL dipole density of
eq.(\ref{density}), our argument {\it does not} assume that the
scattering probability at $R(t)$ is dominated by one pomeron
exchange. The only assumption is, that parametrically the total
unitarized probability is the same as the one pomeron one.
We use the criterion of dipole density only as an indicator
for the magnitude of the total probability. The total scattering probability
is given by
\begin{equation}
  P(b)=\sum_{m=1}^{\infty} \gamma_m(x,r)P_m(x_0,x,b,t)\, ,
 \label{prob} 
\end{equation} 
where $P_m(x_0,x,b,t)$ is the probability to find in the
projectile wave function $m$ dipoles of size $x$ at transverse
coordinate $b$ within the area of the target radius $R_0$ 
and $\gamma_m$ is the probability of the
scattering of an $m$-dipole state. In fact one should also sum
over all dipole sizes smaller than $R_0$. We have neglected this
summation in eq.(\ref{prob}) thus somewhat underestimating the
total probability. Since $\gamma_m\ge\gamma$, the probability is
bounded from below as

\begin{equation}
 P(b)\ge \gamma
 \sum_{m=1}^{\infty} P_m(x_0,x,b,t)\, .
\end{equation}
For dipoles of size $Q_s^{-1}$, the scattering probability $\gamma$ is
of order unity. Thus the only condition that we use is
\begin{equation} 
  \sum_{m=1}^{\infty} P_m(x_0,x,b,t)=O(1)
\end{equation}
 whenever
\begin{equation}
  n(x_0,x,b,t)=\sum_{m=1}^{\infty} mP_m(x_0,x,b,t)=O(1)\ \ . 
\end{equation} 
The only way this condition can be violated, is if the wave function
is dominated (with exponential accuracy !) by the trivial
configuration with no dipoles, even when the average dipole number
is one.
Although the dipole model wave function is known to have
relatively large fluctuations, there is nothing in its known
properties \cite{muellerdip,salam} to suggest such an extreme
behaviour. In fact for the explicit exponential model used in
\cite{muellerdip,salam} our condition clearly holds.

To summarize, in the target rest frame, the violation of unitarity by the BK
evolution can be understood as follows: Start with a single dipole
scattering on the hadronic target of transverse size $R_0$. With
increasing energy the projectile dipole emits additional dipoles
strictly according to the BFKL evolution. The density as well as
the transverse size of the projectile state thus grows. The
increase in density leads to increasing importance of multiple
scatterings which are properly accounted for in the BK derivation.
This ensures that the scattering probability saturates locally. In
the saturation regime, as long as the size of the projectile state
$R(t)$ is smaller than the target size $R_0$, the cross section
grows essentially only due to surface effects,
\begin{equation}
  \sigma = \pi R^2_0+2\pi R_0x_0
  \exp\hspace{-.1cm}\Big [ {\alpha_s N_c\over 2\pi} \epsilon t\Big ]\, .
\end{equation}
As long as ${\alpha_sN_c\over 2\pi} \epsilon t<\ln{R_0\over x_0}$,
the cross section is practically geometrical. However once the
energy is high enough, so that the projectile size is larger than
that of the target, the total cross section is determined by the
former and grows exponentially with rapidity
according to eq. (\ref{radius1}). Thus at rapidities
\begin{equation}
t>{1\over \alpha_sN_c}\ln{R_0\over x_0}\, , \label{tbk}
\end{equation}
the BK evolution is nonunitary and can not be applied.

This also illustrates that the applicability of the BK evolution
crucially depends on the nature of the target. If the target is
thick enough, so that the multiple scatterings become important
before the growth of the projectile radius does,  and if the
target is wide enough, so that saturation occurs before the
projectile radius swells beyond that of the target, then there is
an intermediate regime in which the inelastic cross section
remains practically constant and equal to $\pi R^2_0$. Then BK
applies in this intermediate regime. However, if the target is a
nucleon, neither one of these conditions is satisfied. Thus the
tainted infrared behaviour of the BFKL evolution of the projectile
will show up right away and will invalidate the application of the
BK equation.

\section{The target evolution picture}

The BK equation
(\ref{kov}) is valid only in the leading approximation in
$1/N_c$. Beyond the leading order the evolution for a dipole cross
section does not close, but rather is the first in the infinite
hierarchy of equations. This hierarchy was derived in
\cite{balitsky}. It is useful to consider its interpretation in 
the frame where all the energy resides in the target. In this frame 
further increase in energy leads to growth of the target gluon fields.
The evolution equation governs the change in the distribution of the
gluon fields $A_\mu$ in the wave function of the target.
In the particular gauge  used in
\cite{balitsky}, the largest component of the vector potential is
$A^+$. In this gauge the S-matrix for scattering of a fast
fundamental projectile on the target fields is given by the
unitary eikonal factor $U(x)$
\begin{equation}
  U(x)=P\exp\{i\int dx^-T^aA_a^+(x)\}\, , \label{amp}
\end{equation}
where $T^a$ are the generators of the $SU(N)$ group in the 
fundamental representation. The first in the hierarchy of 
evolution equations derived in \cite{balitsky} is
\begin{eqnarray}
  &&{d\over d t}{\rm Tr}<1-U^\dagger(x)U(y)>={\alpha_s\over
    2\pi^2}\int d^2z{(x-y)^2\over
    (x-z)^2(y-z)^2}\nonumber\\
  &&\times \langle\,  
   N_c{\rm Tr}[U^\dagger(x)U(y)]-{\rm Tr}[U^\dagger(x)U(z)]{\rm
   Tr}[U^\dagger(z)U(y)]\, \rangle \, . \label{bal}
\end{eqnarray}
The averaging in eq. (\ref{bal}) is taken over the ensemble
of field strengths characterizing the target, i.e., 
over the target wave function. In the large $N_c$ limit the averages 
in eq. (\ref{bal}) factorize and one recovers eq.(\ref{kov}) with 
the identification
\begin{equation} 
  N(x,y)={1\over N_c}{\rm Tr}\langle\, 1-U^\dagger(x)U(y)\, \rangle\, .
  \label{scattdensity}
\end{equation}
%
\begin{figure}[h]\epsfxsize=10.7cm
\centerline{\epsfbox{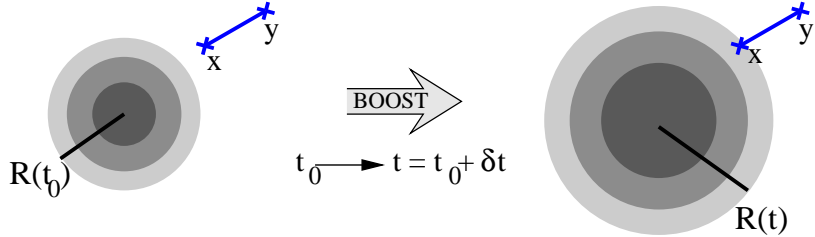}}
\caption{The BK evolution in the target evolution picture.
As long as the dipole leg $y$ is in the white region, only
the evolution of the scattering amplitude $U(x)$ in (\ref{scattdensity})
is non-trivial.
}\label{fig4}
\end{figure}

A physically appealing reformulation of Balitsky's hierarchy of
equations was given by Weigert
\cite{weigert} in terms of a nonlinear stochastic process. The
unitary scattering amplitude $U$ evolves under the action of a
stochastic source
\begin{eqnarray}
  {dU(x)\over dt} &=& g U(x)iT^a \hspace{-.2cm}
     \int \hspace{-.2cm}{d^2z\over \sqrt{4\pi^3}} {(x-z)_i\over (x-z)^2}
     \hspace{-.1cm}
     \left[1-\tilde U^\dagger(x)\tilde
              U(z)\right]^{ab}\hspace{-.3cm} \xi^b_i(z)
     \nonumber \\
     && - \frac{i\alpha_s}{2\pi^2}
     \int d^2z {1\over (x-z)^2}\,
     {\rm Tr}[T^a \tilde U^\dagger(x)\tilde U(z)]\, ,
     \label{wei}
\end{eqnarray}
where $U(x)$ and $\tilde U(x)$ are the unitary matrices (\ref{amp}) in 
the fundamental and adjoint representations, respectively. The white 
noise $\xi$ is characterised by Gaussian local correlations
\begin{equation}
  \langle \xi^a_i(t',z')\xi^b_j(t'',z'')\rangle 
  =\delta^{ab}\delta_{ij}\delta(t'-t'')
  \delta(z'-z'')\, .
\end{equation}
This Langevin equation gives rise to an infinite number of
equations for correlators of $U$ which coincide with those derived
in \cite{balitsky}.

Consider then the Langevin equation formulation, eq. (\ref{wei}).
Assume that at the initial rapidity $t_0$ the target is black
within radius $R_0$. This means that for $|z|<R_0$ the matrix
$U(z)$ fluctuates very strongly so that it covers the whole group
space. Let us concentrate on the point $x$ which is initially
outside of this black region. The matrix $U(x)$ then is close to
unity. Thus there is no correlation between $U(x)$ and $U(z)$, and
the second and third terms on the right hand side of eq. (\ref{wei}) can be set
to zero. This is the random phase approximation introduced in
\cite{weigert} and used later in \cite{el}.
Note that this approximation does not linearize the
evolution. Rather it corresponds to equating the nonlinear
term in eq.(\ref{kov}) to unity for $z$ in the black region.

As the target field
ensemble evolves in rapidity, the radius of the black region
grows. As long as the point $x$ stays outside the black region, we
can approximate the Langevin equation by
\begin{equation}
  {d\over dt}U(x)=-\sqrt{{\alpha_s N_c\over{2\pi^2}}}\int_{|z|<R}
  d^2z{(x-z)_i\over (x-z)^2}\xi_i(z)\, . 
  \label{langevin}
\end{equation}
Here we did not indicate explicitly colour indices, since they are
inessential to the argument. We have also neglected the
contribution to the derivative of $U$ that comes from gluons
originating from the sources outside the black region. Those
contributions speed up the growth of $U$, and so by omitting
them we can only underestimate the rate of growth of the radius of
the black region. The formal solution of equation (\ref{langevin})
is,
\begin{equation}
1-U(x,t)= \sqrt{{\alpha_s N_c\over{2\pi^2}}}\int_{t_0}^{t}
d\tau\int_{|z|<R(\tau)} d^2z{(x-z)_i\over (x-z)^2}\xi_i(z)\, .
\label{langevins}
\end{equation}
Squaring it and averaging over the noise term gives
\begin{equation}
  \langle (1-U(x,t))^2\rangle={\alpha_s N_c\over{2\pi^2}}\int_{t_0}^t
  d\tau\int_{|z|<R(\tau)} d^2z{1\over (x-z)^2}\, . \label{langevinso}
\end{equation}
As long as $x$ is outside the black region and $|x|>R$, we can
approximate the integral on the right hand side by
\begin{equation}
\int_{|z|<R(\tau)} d^2z{1\over (x-z)^2}=\pi{R^2(\tau)\over x^2}\, ,
\label{appr}
\end{equation}
and eq.(\ref{langevinso}) becomes
\begin{equation}
  \langle (1-U(x,t))^2\rangle 
  ={\alpha_s N_c\over{2\pi}}{1\over x^2} \int_{t_0}^t
  d\tau R^2(\tau)\, . \label{langevinsol}
\end{equation}
As the black region grows, eventually it will reach the point
$x$. At this rapidity the matrix $U(x)$ starts fluctuating
with an amplitude of order one. Thus, when $R(t)=|x|$, the left
hand side of eq. (\ref{langevinsol}) becomes a number of order one,
which we call $1/\epsilon$. We thus have an approximate equation
for $R(t)$,
\begin{equation}
  {1\over\epsilon}R^2(t)={\alpha_s N_c\over{2\pi}} \int_{t_0}^t
  d\tau R^2(\tau)\, ,
\end{equation}
or in the differential form
\begin{equation}
{d\over dt}R(t)={\alpha_s N_c\over{4\pi}}\epsilon  R(t)\, .
\label{radius}
\end{equation}
At large rapidities therefore the radius of the black region is
exponentially large
\begin{equation}
R(t)=R(t_0)\exp\{{\alpha_s N_c\over{4\pi}}\epsilon (t-t_0)\}\, .
\label{radiuss}
\end{equation}

We thus recover the result of the previous section.

The approximations leading to eq. (\ref{radius}) are not
strictly speaking valid when the point $x$ is on the boundary of
the black region. First, eq. (\ref{appr}) is an underestimate of the
integral, since the inequality $|x|\gg R$ no longer holds. However,
this approximation can lead only to
an underestimate of the rate of growth of $R$. Second, not for all
points $z$ in the black region the term $U(x)U^\dagger(z)$ in
eq.(\ref{wei}) can be dropped. This however is also unimportant,
since when $x$ is on the boundary of the black region, although
the factor $(1-U(x)U^\dagger(z))$ is not striclty unity, it is
still of order one for all points $z$. It is in fact different from unity
only for points $z$ in the vicinity of $x$. Thus, although we can not
determine the exact numerical value of $\epsilon$, the functional
form of the solution as well as its parametric dependence is given
correctly by eq. (\ref{radiuss}).

From the point of view of the evolution of target fields the
violation of the unitarity can be interpreted as follows. The RHS
of eq. (\ref{wei}) is nothing but the total Coulomb
(Weizs\"acker-Williams) field at point $x$ due to the colour charge
sources at points $z$. The stochastic noise tells us that these
colour sources are uncorrelated both in the transverse
plane and in rapidity. For such random sources the square of the
total colour charge is proportional to the area, and this is
precisely the factor $R^2$ in eq. (\ref{langevinsol}). The incoming
dipole thus scatters on the Coulomb field created by a large
incoherent colour charge. Because the Coulomb field is long range,
the whole bulk of the region populated by the sources contributes
to the evolution and leads to rapid growth of $R$. If the field
created by the sources was screened by some mass, the evolution
would be perfectly unitary. To illustrate this point, let us
substitute the Coulomb field $(x-z)_i/(x-z)^2$ in eq. (\ref{wei}) by
an exponentially decaying field $m\exp\{-m|x-z|\}$. It is
straightforward to perform now the same analysis as before.
Eq. (\ref{appr}) is now replaced by
\begin{equation}
  \int_{|z|<R(\tau)} d^2z m^2\exp\{-m|x-z|\}=\exp\{-m|x-R|\}\, .
\label{appr1}
\end{equation}
This leads to the substitution $R^2\rightarrow \exp\{mR\}$ in all
subsequent equations with the end result that
\begin{equation}
R(t)=\alpha_s{\epsilon\over m}t\, ,
\end{equation}
which in fact saturates the Froissart bound.

Thus the reason for the violation of unitarity is that the
evolution is driven by the emission of the long range Coulomb
field from a large number of {\it incoherent} colour sources in
the target.

To conclude this section we would like to discuss the relation of
our results with numerical studies of the nonlinear QCD evolution.
Studies within the framework of the dipole model were reported
in \cite{salam}.  Ref. \cite{salam} does not deal directly with
the nonlinear BK equations, but rather with the onium-onium
scattering in the framework of the dipole model. However as is
clear from our discussion at asymptotically high energies this
distinction is irrelevant. The growth of the transverse size of
the projectile eventually determines the behaviour of the cross
section irrespective of the nature of the target. Indeed our
results are in agreement with those of \cite{salam}. The numerical
results of \cite{salam} clearly indicate, that even though the
scattering probability is unitarized locally in the impact parameter
space, the total cross section keeps on growing exponentially with
$t$ (Figs. 9 and 10 of \cite{salam}). From Fig. 10
of \cite{salam} we conclude that the power of the exponential
is about $.75\omega$, where $\omega$ is the leading BFKL
exponential. Interestingly, our rough estimate (\ref{epsilon1}) 
is in reasonable qualitative agreement with this numerical result.

Our results eqs.(\ref{radius1},\ref{radius})
are in apparent contradiction with the conclusions of numerical
work \cite{lt,braun}. The origin of this discrepancy is
that these references solve eq. (\ref{kov})
within the local approximation, assuming that important
contributions come only from the dipole sizes which are smaller
than the impact parameter. Within this approximation the
dependence on the impact parameter in eq. (\ref{kov}) becomes
parametric, and the growth of the total cross section is
determined entirely by the shape of the initial condition. The
Froissart bound is then saturated for the exponential initial
profile of $N(b)$. The physics here is simple. In the local
approximation of \cite{lt,braun} the gluon density of the target
(and the scattering probability $N$) evolve at all impact
parameters $b$ according to the same exact translationally
invariant equation. The density locally grows at all impact
parameters at the same rate. This rate is the same as the growth
of the saturation momentum and is power like with rapidity,
eq.(\ref{qs}). Thus if one starts from initial configuration with
the exponential density profile $g(b)=\exp\{-b/R_0\}$, after the evolution
to rapidity $t$ it becomes
\begin{equation}
g(b,t)=\exp\{-b/R_0+\alpha_s\lambda t\}\, .
\end{equation}
The scattering probability on such a system is unity at impact
parameters for which $g(b)\ge 1$. Thus the highest impact
parameter that contributes to the total cross section at rapidity
$t$ is $b_{max}={\alpha_s\lambda\over R_0}t$, and the cross
section\footnote{In this discussion we neglect the dependence of
$N$ on the size of the dipole. Strictly speaking such dependence
is of course present, and it determines the rapidity at which the
asymptotic behaviour of $\sigma$  sets in. The 
asymptotic form of $\sigma$  however is independent on the dipole size.} is
$\sigma=\pi b^2_{max}\propto t^2$ . For initial Gaussian
distribution on the other hand, the same argument leads to the
linear growth of the cross section with $t$.

This feature, namely that the asymptotic form of $\sigma$ is
determined by the initial distribution is clearly an artifact of
the local approximation. The reason the local approximation leads
to this behaviour, is that it neglects the effects of far away
black regions (where $N=1$) on the scattering probability in the
grey areas (where $N<1$). As is apparent from our analysis in
eq.(\ref{langevin}-\ref{langevinsol}), it is precisely the effect
of the far away black regions that drives the growth of the total
cross section. This is due to the long range Coulomb fields
originating in the central black region. In fact, the only
contributions we kept on the RHS of eq.(\ref{langevin}) are due to
dipoles with sizes of the order of the impact parameter. In this
respect our discussion is orthogonal to that of \cite{lt,braun}.
It is clear from the comparison of our results to those of
\cite{lt,braun}, that the effect of these long range fields on the
total cross section is far greater than that of the local
translationally invariant part of the evolution. Even if
one starts from an exponential density profile, the full BK
evolution generates {\it power like} and not exponential tails
in the density at large $t$. These power like tails dominate the
total cross section and lead to the universal exponential growth of
$\sigma$ with rapidity. The local approximation is adequate for
studying the behaviour of $Q_s(t)$ in the dense central region, as
this is determined by local effects. It is however not a good
approximation for the total cross section, which is dominated by
the evolution of long range Coulomb fields.

\section{One Pomeron, two Pomeron; hard Pomeron, soft Pomeron}

The linear evolution of the projectile wave function inherent in 
the BK evolution
is a direct consequence of the large $N_c$ limit. In this limit
one can neglect the interactions between the dipoles in the
projectile wave function. Individual Feynman diagrams which
contain dipole-dipole interactions are suppressed by powers of
$1/N_c$. Thus, the dipoles in the projectile wave function do not
interact during the evolution and they scatter independently of
each other.
However, the number of interacting diagrams grows very fast with the
number of interacting dipoles. As the number of dipoles in the
wave function which can interact with each other becomes 
$O(N)$, the number of the suppressed diagrams becomes $O(N^2)$ and
the suppression disappears.  At high enough rapidity, where
the dipole-dipole interactions are important, the evolution
equation (\ref{kov}) breaks down and wave function saturation
effects start to play an essential role. Although equation 
(\ref{bal}) contains some $1/N_c$ corrections, those are only "group
theoretical" corrections reflecting the fact that at finite $N_c$
a gluon is not strictly equivalent to a $\bar q q$ pair. It
does not contain the corrections due to the interactions of
gluons (or dipoles) in the projectile wave function.

The interaction probability of a dipole of size $r$ with
another dipole of similar size in the projectile wave function is
of the order $\alpha_sr^2n(r)$, while the probability of the
direct interaction with the target is $\gamma(r)$. The multiple
scattering corrections are thus more important as long as
$\alpha_sr^2n(r)\ll \gamma(r)$. The projectile wave function
corrections become important as the density grows so that
\begin{equation}
   r^2n(r)=\exp\{\omega t\}={\gamma\over \alpha_s}\, .
  \label{nc}
\end{equation}
 Assuming that for a large
target $\gamma$ is of order one, this happens for rapidities
\begin{equation} 
  t\propto{1\over \omega}\ln 1/\alpha_s 
  \sim {1\over (N_c\, \alpha_s)}\ln 1/\alpha_s\, .
  \label{td}
\end{equation}
The "subleading" large $N_c$ nature of these corrections is clear
if one traces back the explicit $N_c$ dependence of (\ref{td}). 
In fact, the coupling constant $\alpha_s$ occurs in $n(r)$ only 
in the combination $\alpha_s\, N_c$, since the BFKL evolution 
neglects dipole-dipole interactions. Thus the $\alpha_s$ in the 
numerator of eq.(\ref{nc}) is explicitly subleading in $N_c$ and 
indeed accounts for dipole-dipole interactions.

The above illustrates that the BK resummation
improves on the BFKL evolution when the target is dense, that is
the scattering probability $\gamma$ is larger than order
$\alpha_s$. On a large nucleus of atomic number $A$ one expects
$\gamma\propto\alpha_sA^{1/3}$, and thus in this case BK resumms
all corrections in powers of $\alpha_sA^{1/3}$ \cite{kovchegov}.
However even in this case the validity of the BK equation remains
limited.
The scattering probability $\gamma$ is only large at central
impact parameters. At peripheral impact parameters the density in
the target drops to zero, and so for the peripheral scattering
events the applicability of BK is no better than that of BFKL. In
particular, the diffusion contributions play the same role, since they
are not associated with high partonic densities. Since at very high
energies the total cross section is dominated by peripheral
events, one does not expect the BK evolution to be a valid
approximation in calculating the total cross section. Comparing
eqs.(\ref{tbk}) and (\ref{td}) we conclude that for large targets,
at rapidities at which the total cross section is dominated by 
peripheral impact parameters, the wave function saturation
effects are important even for small projectiles. Thus even within
perturbation theory, the wave function saturation
effects are bound to be the ones that determine the growth of the total
cross section with energy.

Could wave function saturation effects lead to a perturbative
unitarization of the total cross section?  As seen above, the
violation of the Froissart bound in the BK approximation is due to
large and incoherent colour charge fluctuations in the black
region. If there was a mechanism to ensure strong correlations
such that the total colour charge in a region of fixed size $L$ is
zero, then the incoming dipole would feel the Coulomb field only
within the fixed distance $L$ from the black region. Thus the new
charges produced by the evolution would only "split off" the edges
of the black region rather than from its bulk. This scenario is
equivalent to exponential decay of the field, and would lead to a
unitary evolution, as expected for a confining theory like QCD.
However, it does not seem plausible that this type of
correlations can be enforced on the system by a perturbative
mechanism, nor that perturbation theory can generate any kind of 
"mass" for the gluons which would lead to exponential decay of 
gluon fields generated by sources in the black region. We thus
expect that any
perturbative corrections will generate power like tails of gluon
density at large distances from the black region. The growth of
these power tails with rapidity will inevitably lead to an exponential
growth of the cross section.

We expect, however, that wave function saturation effects diminish
the exponent $\epsilon$ significantly relative to the BK value
${\alpha_sN_c\over 2\pi}\epsilon=.75\omega$. The evolution equations
derived in~\cite{jklw} go beyond the BK equation by including 
wave function saturation effects. They provide a
well-defined perturbative framework for calculating this effect.

We have seen in the previous sections that QCD perturbation theory
predicts the existence of two distinct physical  mechanisms
for the power growth of hadronic cross sections with energy. One
is due to the fast growth of the partonic densities, while the
other is due to expansion of hadronic states in the transverse
plane. The first mechanism is the leading one for systems
containing a small number of partons. When the partonic density
reaches the critical value of order $1/\alpha_s$, further growth
of the density is cut off by the perturbative saturation effects.
The transverse expansion mechanism is however likely to survive in
this situation, and thus it should become the leading driving
force for the perturbative growth of cross sections in dense
systems.

The first mechanism with the perturbative BFKL exponent of around
$.3-.4$ fits very nicely with the so called hard Pomeron utilized to
fit the hadronic cross sections of small systems \cite{twopom}.
The exponent due to the transverse expansion has not been
calculated yet, although the equations which determine it
have been derived in \cite{jklw}. Although we do not know
yet the value of this exponent, we know for certain, that
it is smaller than the BFKL one. One is naturally
lead to ask whether this perturbative growth of the transverse
size of saturated systems underlies the experimentally observed
power growth of the cross sections in purely hadronic processes,
like $pp$ and $\bar pp$.
This type of expansion, and with it the power like growth of the cross
section should cease at asymptotically high energies, where the 
perturbation theory must become invalid.

This picture suggests that
nonperturbative effects are called for only in order to
unitarize the cross section at asymptotically high energies, but
not in order to furnish the mechanism for its fast growth in the
pre-asymptotic regime. This is also  natural from the
following perspective. The Froissart-like behaviour is associated
with the existence of a gap in the spectrum, as can be illustrated
by a simple and
intuitive argument due to Heisenberg \cite{heisenberg}.
In a theory with a mass gap, the profile of the distribution of
matter density in any target must decay exponentially at the 
periphery, $\rho(b)\propto \exp\{-mb\}$. As this target is struck
by a projectile, in order to produce an inelastic scattering event
at least one particle must be produced. Assuming that the
scattering is local in the impact parameter plane, the region of
the overlap of the probe and the target must therefore contain
energy at least equal to the mass of the lightest particle, $m$.
For scattering at energy $E=s/m$ in the frame where all the energy
resides in the target, the target energy density is $E\rho(b)$.
Thus the scattering can only take place
for impact parameters smaller than those that satisfy
$E\exp\{-mb\}=m$. Thus $b_{max}={1\over m}\ln\{s/m^2\}$, which is
equivalent to the Froissart bound. Conversely, if the
cross section grows as a power of energy, then the
density distribution in the target is not exponential but 
power like. With $\rho(b)\propto b^{-\lambda}$ one obtains
$b_{max}\propto s^{1\over\lambda}$. Since the power growth of
hadronic cross sections persists in a large interval of energies,
one expects that for a range of impact parameters the density
distribution in hadronic states is power like. Perturbation
theory provides a natural explanation of such a power
like distribution. The tails of perturbative distribution are due
to massless gluon fields emitted from the colour charges in the
target. Even though the target is neutral, it always possesses a
multipole moment of some order, and thus perturbatively is always
accompanied by a long range power like tail of massless gluon
field.
%
\begin{figure}[h]\epsfxsize=10.7cm
\centerline{\epsfbox{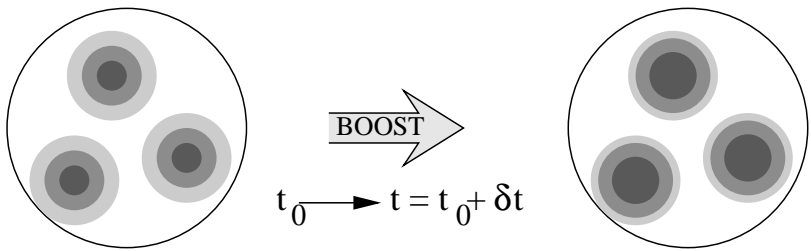}}
\caption{Schematic picture of the proton as a loosely bound system 
of three constituent quarks which provide ``black'' building blocks
at initial rapidity $t_0$. Under boost, the blackness of these
regions does not change since the gluon density is saturated, but 
the transverse size of the black region grows.
}\label{fig5}
\end{figure}
%
The precondition for applicability of this perturbative mechanism
in hadronic systems is that hadrons themselves are built from small
"black" building blocks. As explained earlier, the gluon fields
at periphery are emitted from the bulk of the black disk, and not from its
boundary. Thus, in order to explain the preasymptotic power like rise
of hadronic cross sections by perturbative Coulomb like gluon fields,
the radius of these black region must be smaller than the
confining scale of QCD.
It is in fact widely believed that QCD does naturally contain the scale of the
right order - the scale associated with chiral symmetry breaking.
In particular the radius of constituent quark is believed to be
at most $.3$ fm and perhaps even smaller\cite{nyiri}.
This supports the phenomenological picture of 
the proton as a loosely 
bound system of three small constituent quarks. The 
nonperturbative confining force keeps these quarks confined within the 
spatial region of radius $.8-.9$ fm.

A quantitative analysis of the non-linear evolution 
equations~\cite{jklw} which could substantiate our proposal of 
the perturbative soft pomeron, is not available yet. There are 
however some qualitative consequences of the above picture that 
can be compared to experimental data. Let us check first whether 
the value of the cross section in our model is in rough agreement
with experiment. The total cross section of $pp$ scattering at the
lowest energy where the Reggeon contributions are not important,
$\sqrt{s}\propto 50$ GeV is $45$ mb. The total cross section
for the scattering of two black disks of diameter $d$ is $2\pi d^2$.
Thus in the simplest model in which the proton is  composed 
of three completely ``black'' constituent quarks of diameter $d$, 
the total $pp$ cross section
is $3\times 3 \times  2
\pi d^2$. Equating this to $45 mb$ we find $d=.28 $ fm. 
This should be considered as a lower bound on the value of $d$.
It is more likely that the constituent quarks are only black in
the center and have grey peripheral regions, see  Fig.~\ref{fig5}. 
For peripheral $qq$  events the scattering occurs with
probability  $f<1$. Incorporating this roughly 
as an average ``greyness'' factor 
in the formula for the cross section gives $d=.28f^{-1/2}$ fm. 
For $f=.5$ we have $d=.4$ fm.
We may thus think of the proton as a
collection of three loosely bound constituent quarks each described
by a disk of $d = .3-.4$ fm which is essentially black in its center
but grey at its boundary, see Fig.~\ref{fig5}. 

This picture is quite
remarkable, since the "active" area inside the proton in our model
is much smaller than the proton radius and thus one may have
worried that the model will underestimate the total cross section.
This however does not happen. Further support for this picture
comes from the ratio ${\sigma_{\pi p}\over \sigma{pp}}$ which is
very close to $2/3$.
The pomeron contributions to the total hadronic cross sections 
are parametrized as $\sigma_{pp}=21.70 s^{0.808}$ and 
$\sigma_{\pi p}=13.63 s^{0.808}$ \cite{dewolf}.
This is consistent with $\pi$ having two
constituent quarks.

Another feature of the model is that although the quarks are
black, the proton itself is not. Thus we expect the ratio of the
elastic to total cross section to be well below the black disk
value of $1/2$. The experimental value of this ratio at 
$\sqrt{s}=50$ Gev is indeed just below $1/5$ \cite{dewolf}.

Another global characteristics of the scattering is the radius of
the proton as measured via the shrinkage of the elastic peak,
${d\sigma_{el}\over dt}\propto \exp \{{R^2\over 4}t\}$. For the
elastic $pp$ scattering this gives the value of $R$ consistent
with the proton radius $R=.8$ fm \cite{dewolf}. Again at first
sight this sounds like trouble, since it is much larger than the
radius of a constituent quark. However one should realize that the
radius of the individual quark is not relevant for this particular
quantity. The process is elastic if the proton as a whole emerges
from the interaction intact. Thus indeed it is the radius of the
proton and not of a quark that should determine the $t$-dependence
of the elastic cross section in our model. The radius of the quark
may emerge in a similar way in the processes "elastic" with
respect to individual constituent quark scattering. Whether one
can define a subset of final states that correspond to such a
process is an interesting question, but at present it is not clear
to us how to do it.

The picture of the proton
as built from three small constituent quark
was envoked in the nonperturbative model for the high energy scattering in
\cite{kopelovich}.
The physics of \cite{kopelovich} is however quite different from that of 
our proposal.
The quarks themselves in \cite{kopelovich} are not thought of as being
 black, and
the growth of the quark-quark cross section is due to the increase 
of the density of the
gluon cloud surrounding an individual quark, 
rather than to the increase in its transverse size.
Ref. \cite{dima} also appeals to the scale of $.3$ fermi although
not explicitly in connection with the size of constituent quarks. Again, 
however the mechanism of the growth of the cross section in \cite{dima} 
appears to be the same
as the leading BFKL mechanism, that is the growth of gluonic density. The
role of instanton effects in \cite{dima}
is to limit the gluon emissions only to within
the transverse sizes smaller than $.3$ fermi, and thus to cut off the
expansion in the transverse plane. In contrast our picture assumes that
nonperturbative effects (perhaps instantons) are responsible for the buildup
of a black gluon cloud around each quark at low energy. The subsequent
evolution in energy is dominated by perturbative swelling of these black 
regions. Eventually, when the size of the quarks reaches confining scale, 
other nonperturbative effects kick in and cut off further power like 
growth of the cross section. The physics of these nonperturbative effects 
is presumably the physics of confinement.

We note also that the two distinct physical mechanisms for two pomerons
have a consequence that they will appear with different probability
in different processes. For example in $pp$ scattering there are rare proton 
configurations which do not have the typical hadronic structure, but 
contain quarks closely bunched together in coordinate space. 
Those are the configurations responsible for the colour transparency 
effects in the hadronic scattering \cite{transp}. These 
configurations are rare and short lived. Thus they do not have time to 
develop the gluon clouds around the quarks that could make them "black". 
These configurations will predominantly evolve towards the increase in 
density, and thus will have energy dependence of the hard pomeron.
This is consistent with the findings of \cite{twopom} that hard pomeron
is present already in purely hadronic processes.

On the other hand one does not necessarily 
expect the soft pomeron to appear in 
DIS even at low $Q^2$. The reason is that even though 
at low $Q^2$ the DIS cross section has large contributions from the photon 
fluctuations into the states of hadronic size, these states do not 
live long enough on hadronic time scale, and thus do not have time to
develop dense gluon clouds around the quarks. The energy dependence of such 
large but dilute states would then be of the hard pomeron nature.
This appears to be consistent with the recent results which do not require 
the soft pomeron to fit DIS data even at very small $Q^2$ \cite{gluonss}.

Perhaps the most appealing feature of this scenario is that it gives hope
to understand the soft pomeron within the well defined, {\it bona fide}
perturbative framework. The equations that resum the wave function 
saturation effects has been derived in \cite{jklw}.
Even though their numerical study is probably much more involved
than that of eqs.(\ref{kov},\ref{bal}), we think the question is
interesting enough to motivate such an undertaking.

{\bf Acknowledgements} This work has been supported in part by
PPARC. U.A.W. thanks the Department of Mathematics and Statistics,
University of Plymouth for hospitality while part of this work was
being done. We are indebted to J.G. Milhano and H Weigert for interesting
discussions during the early stages of this work. We also
thank M. Braun, Yu. Kovchegov, E. Levin and K. Tuchin
for useful discussions and correspondence.

\pagebreak


\begin{thebibliography}{99}
\bibitem{expon} More accurately one should require that the number of
dipoles of size $Q_s^{-1}(t_0)$ in the area of the overlap
with the target ($\pi R_0^2$) is at least one. This refinement however
only modifies the pre exponential factor in eq.(\ref{radius1}),
the calculation of which is in any case beyond the scope of this paper.
To determine this pre exponential factor one would have to treat more
carefully the pre exponential factors in eq.(\ref{density}).
Eq.(\ref{radius1}) is thus valid only with exponential accuracy.
\end{thebibliography}
\end{document}